%%%%%%%%%%%%%%%%%%%%%%%%%%%%%%%%%%%%%%%%%%%%%%%%%%%%%%%%%%%%%%%%%%%%%%%%%%%%%%%
%% Page layout:
\magnification = \magstep 1
\baselineskip = 16truept
\vsize = 9.05truein
%\voffset = -0.05truein
\nopagenumbers
\raggedbottom
\headline = {\tenrm\ifnum \pageno > 1 \centerline{-- \folio\ --}\else
\centerline{To appear in {\it Publications of the Astronomical Society of
the Pacific}}\fi}
\def\sec#1 #2{\bigskip\medskip\centerline{\apj {#1}~{#2}}}
\def\subsec#1 #2{\bigskip\centerline{{#1}~{\sl #2}}\medskip}
\def\subsubsec#1 #2{\medskip\centerline{{#1}~{#2}}\medskip}
\def\fpsec{\bigskip}

%% Fonts:

 at 12truept

\font\smrm = cmr9

\font\apj = cmcsc10

%% Definitions:
\def\et{et al.\ }
\def\eg{e.g.,\ }
\def\ie{{i.e.,\ }}
\def\hi {\noindent \hangindent=2.5em}

\def\E#1{$10^{#1}\, {\rm ergs\> s^{-1}}$}
\def\EE#1 #2{$#1 \times 10^{#2}\, {\rm ergs\> s^{-1}}$}
\def\FF#1 #2{$#1 \times 10^{#2}\, {\rm ergs\> cm^{-2}\> s^{-1}}$}

\def\Hb{H$\beta $}

\def\HeII {He~{\apj ii}}

\def\Msun{\ifmmode M_{\odot} \else $M_{\odot}$\fi}

\def\errora#1 #2 #3{$#1^{\scriptscriptstyle
+#2}\!\!\!\!\!\!\!\!\!_{\scriptscriptstyle -#3}$}
\def\errorb#1 #2 #3{$#1^{\scriptscriptstyle
+#2}\!\!\!\!\!\!\!\!\!\!\!_{\scriptscriptstyle -#3}$}
\def\errorex#1 #2 #3 #4{$#1^{\scriptscriptstyle
+#2}\!\!\!\!\!\!\!\!\!\!\!_{\scriptscriptstyle -#3} \times 10^{#4}$}

\def\farcs{\hbox{$.\!\!^{\prime\prime}$}}
\def\farcm{\hbox{$.\!\!^{\prime}$}}
\def\hr{\hbox{$^{\rm h}$}}
\def\mn{\hbox{$^{\rm m}$}}
\def\fs{\hbox{$.\!\!^{\rm s}$}}

%% Inputs:
\input epsf

%%%%%%%%%%%%%%%%%%%%%%%%%%%%%%%%%%%%%%%%%%%%%%%%%%%%%%%%%%%%%%%%%%%%%%%%%%%%%%%

\topglue 0.5truein

\centerline{THE NUCLEAR SPECTRAL ENERGY DISTRIBUTION OF NGC 4395,}
\centerline{THE LEAST LUMINOUS TYPE 1 SEYFERT GALAXY}

\bigskip\bigskip
\centerline{\apj Edward C.\ Moran$^{1,2}$, Alexei V.\ Filippenko$^2$, Luis C.\ Ho$^3$,}

\bigskip
\centerline{\apj Joseph C.\ Shields$^4$, Tomaso Belloni$^5$, Andrea Comastri$^6$,}

\bigskip
\centerline{\apj Steven L.\ Snowden$^7$, and Richard A.\ Sramek$^8$}

\bigskip\bigskip\medskip
\centerline{ABSTRACT}
\bigskip
We present X-ray ({\sl ROSAT}), infrared, and radio observations of NGC~4395,
which harbors the optically least luminous type~1 Seyfert nucleus discovered
thus far.  In combination with published optical and ultraviolet spectra,
we have used these data to assemble the broadband spectral energy
distribution (SED) of the galaxy's nucleus.  Interestingly, the SED of
NGC~4395 differs markedly from the SEDs of both quasars and typical
low-luminosity active galactic nuclei, which may be a manifestation of
the different physical conditions (\ie black hole masses, accretion rates,
and/or accretion modes) that exist in these objects.  The nuclear X-ray
source in NGC~4395 is variable and has an observed luminosity of just
$\sim$~\E {38}.  Although this emission could plausibly be associated
with either a weak active nucleus or a bright stellar-mass binary system,
the optical and ultraviolet emission-line properties of the nucleus
strongly suggest that the X-rays arise from a classical AGN.

\medskip\noindent {\it Subject headings:} galaxies: individual (NGC 4395) ---
galaxies: Seyfert --- X-rays: galaxies

\vfill

{\parindent 1em
\vfootnote {$^1$} {Chandra Fellow.\vskip -14truept}

\vfootnote {$^2$} {Department of Astronomy, University of California,
Berkeley, CA 94720-3411.\vskip -14truept}

\vfootnote {$^3$} {Carnegie Observatories, 813 Santa Barbara St., Pasadena,
CA 91101-1292.\vskip -14truept}

\vfootnote {$^4$} {Physics and Astronomy Department, Ohio University, Athens,
Ohio 45701-2979.\vskip -14truept}

\vfootnote {$^5$} {University of Amsterdam, Kruislaan 403, NL-1098
Amsterdam, The Netherlands.\vskip -14truept}

\vfootnote {$^6$} {Osservatorio Astronomico di Bologna, via Zamboni 33,
I-40126, Bologna, Italy.\vskip -14truept}

\vfootnote {$^7$} {Code 662, NASA Goddard Space Flight Center, Greenbelt,
MD 20771.\vskip -14truept}

\vfootnote {$^8$} {NRAO, P.O.\ Box O, Socorro, NM 87801.}

\par}

\break
\sec {1.} {Introduction}

\fpsec
Toward the conclusion of a large survey aimed at quantifying the faint end
of the luminosity function of active galactic nuclei (AGNs), Filippenko \&
Sargent (1989) discovered an extremely low-luminosity type~1 Seyfert nucleus
in the nearby dwarf galaxy NGC~4395.  With an absolute $B$ magnitude of
about --10, the AGN in NGC~4395 is some $10^8$ times weaker than the most
luminous known quasars.  Nevertheless, spectroscopy in the optical and
ultraviolet bands has indicated that this feeble object is similar to classical
Seyfert~1s and QSOs in many respects (Filippenko \& Sargent 1989; Filippenko,
Ho, \& Sargent 1993; Filippenko \& Ho 1999).  For example, the equivalent
widths of the broad optical emission lines are typical of type~1 Seyferts,
and the optical/UV continuum is nearly featureless, showing little evidence
for stellar absorption lines.  In addition, the narrow emission lines span a
wide range of ionization from [O~{\apj i}] to [Fe~{\apj x}]; as in other
AGNs, these lines probably arise from gas that has been photoionized by a
reasonably hard nonstellar continuum (Kraemer \et 1999).

In this paper we re-examine NGC~4395 in the context of its more luminous
counterparts through a comparison of their radio--to--X-ray spectral energy
distributions (SEDs).  The SEDs of AGNs have, in principle, the power to
reveal detailed information about the accretion processes occurring within
them.  For quasars, which can dominate the total output of their host
galaxies at all wavelengths, measurement of the SED is usually
straightforward.  The SEDs of lower luminosity AGNs, however, can be
significantly compromised by contamination from extranuclear sources.  In
order to minimize this contamination, the SEDs of these objects should be
constructed from observations of the highest possible angular resolution
(\eg Ho 1999a).  Bearing this in mind, we have assembled the nuclear SED
of NGC~4395 between 1.4~GHz and 2~keV using data that have, at most
wavelengths, arcsecond-scale resolution.  Direct comparison of this SED
with the SEDs of quasars and typical low-luminosity AGNs offers valuable
insight into the nature of the nuclear activity in NGC~4395.

\sec {2.} {Multiwavelength Data}

\subsec {2.1.} {Soft X-Ray Observations}

NGC~4395 was the target of a 17050~s observation in the 0.1--2.4~keV energy
range with the Position Sensitive Proportional Counter (PSPC) on board
{\sl ROSAT}.  A description of the satellite and X-ray detector are given by
Tr\"umper (1983) and Pfeffermann \et (1986).  The observation produced two
datasets, an 8007~s exposure collected between 1992 June 2 and June 11 UT,
and a 9043~s exposure collected between 1992 June 17 and June 19 UT.  As
Figure~1 indicates, five X-ray sources were detected within $3'$ of the nucleus
of NGC~4395. One of them (source~A), at $\alpha_{2000}$ = 12\hr 25\mn 48\fs6,
$\delta_{2000}$ = +33$^\circ$ $32'$ $51''$, is centered just $4''$ from
the optical position of the nucleus (McCall, Rybski, \& Shields 1985),
well within the 1~$\sigma$ position error radius for on-axis PSPC sources.
The radial profile of source~A is consistent with that of a point source.
None of the other X-ray sources, including the weak source $48''$ south of
the nucleus (source B), corresponds with an optical feature in the galaxy.

NGC~4395 was also observed with the {\sl ROSAT\/} High Resolution Imager
(HRI) for 11,353~s on 23--24 June 1996 UT.  With $\sim$~5$''$ angular
resolution, the HRI allows source positions and spatial extents to be
measured with greater accuracy than can be achieved with the $\sim$~30$''$
resolution afforded by the PSPC.  Unfortunately, the nuclear X-ray source
in NGC~4395 was not detected in the HRI observation.  We have, however,
used the HRI data to evaluate the astrometric quality of the PSPC image.
Comparing the HRI image to the Palomar Observatory Sky Survey, we find
four HRI sources to have optical counterparts within $3''$--$9''$.  Each
of these sources was detected in the PSPC image at positions that closely
match their HRI positions.  Thus, we are confident that source~A in Figure~1
is indeed coincident with the nucleus of NGC~4395.

For spectral and temporal analysis of source~A, we extracted PSPC counts
within a $38''$ radius circle, which we offset a few arcseconds to the
north of the nucleus of NGC~4395 to minimize the possible contamination from
source~B.  A total of 157 counts were detected in this region.  Using the
documented characteristics of the {\sl ROSAT\/} PSPC on-axis point-spread
function (Hasinger \et 1992), we calculate that no more than 2 of these counts
could arise from source~B, assuming it is spatially unresolved.  To estimate
the sky background, we collected counts within an annulus centered on the
nucleus with inner and outer radii of $4'$ and 5\farcm25, respectively.
This region excludes the other sources shown in Figure~1, but is close
enough to the optical axis that we may neglect vignetting effects.  Scaling
the number of counts in the annulus to the area of the source region implies
a background level of 37 counts in the source region.  Thus, the significance
of the detection of the nucleus of NGC~4395 is 9.6~$\sigma$, and the net count
rate of the source, averaged over the entire observation, is ($7.0 \pm 0.7$)
$\times$ $10^{-3}$~counts~s$^{-1}$.  For spectral fitting, we rebinned the
PSPC spectrum to have a minimum of 20 counts (source plus background) per
channel.  With the exception of the lowest energy channel, each spectral
bin has a signal-to-noise ratio of at least 3.

An independent analysis of the {\sl ROSAT\/} data for NGC~4395 has been
carried out recently by Lira \et (1999).  Although some of their findings
are qualitatively similar to those we describe below, our analysis methods
and quantitative results differ in detail.  In particular, the PSPC count
rate derived by Lira \et is higher than ours, probably due to contamination
by Source~B in their large extraction aperture.

\subsubsec {2.1.1.} {The Soft X-ray Spectrum}

With just 120 source photons, only a limited investigation of the soft X-ray
spectral characteristics of NGC~4395 is possible.  Therefore, we have applied
only simple, single-component models to the PSPC spectrum.  An absorbed
power-law model ($F_{E} \propto E^{-\alpha}$), which is frequently employed
in AGN studies, yields the following best-fit spectral parameters: a very
flat power-law energy index of $\alpha \approx -0.1$, and an absorption
column density of $N_{\rm H} = 1.6 \times 10^{20}$ cm$^{-2}$, which is
consistent with the value of the Galactic neutral hydrogen column in the
direction of NGC~4395 ($1.43 \times 10^{20}$ cm$^{-2}$; Murphy \et 1996).
We were surprised to find that, despite the limited statistics of the source
counts, the fit afforded by the power-law model is formally unacceptable;
with $N_{\rm H}$ fixed at the Galactic column density, $\chi^2 = 10.9$ for
4 degrees of freedom.  Very similar results were obtained using a different
background region, so we are confident that the poorness of the power-law
fit is not an artifact of the background subtraction.  Fits involving
single-component thermal models or an intrinsically steep, but heavily
absorbed, power law proved to be inadequate as well.  As Figure~2 illustrates,
the residuals from the best power-law fit exhibit a systematic S-shaped wiggle,
which may signify the presence of complex features (\ie multiple absorption
or emission components) in the soft X-ray spectrum.  Unfortunately, the
quality of the PSPC data is insufficient for a direct investigation of this
possibility.  The power-law fit is, however, adequate for the construction
of the nuclear SED of NGC~4395, and we use it below for that purpose.

\subsubsec {2.1.2.} {X-Ray Variability}

As noted above, the PSPC observation of NGC~4395 was divided into two segments
approximately equal in exposure and spaced by about 12~days.  Although the
overall statistics are too poor to allow the production of a proper light
curve, we have looked for possible variability of the nuclear X-ray source
between these segments.  To do this, we extracted source and background counts
for each dataset individually using the procedure described above.  We find
that while the background did not vary between the two observations, the
nucleus of NGC~4395 did.  We measure a source count rate for the first interval
(8007~s) of ($3.7 \pm 0.9$) $\times$ $10^{-3}$ counts~s$^{-1}$.  During the
second interval (9043~s), the source count rate increased to ($10.0 \pm 1.2$)
$\times$ $10^{-3}$ counts~s$^{-1}$.  Thus, the nucleus of NGC~4395 varied by
a factor of $\sim$~3 (significant at the 4.3~$\sigma$ level) on this short
time scale.  For comparison, we examined other sources in the field
for variability; the count rate of source~E in Figure~1 increased by 30\%
between the two observations (2.7~$\sigma$ significance), while that of a
source $\sim$~$8'$ to the southwest of NGC~4395 (not shown in Fig.~1)
decreased by 12\% (0.8~$\sigma$ significance).

Variability explains why the nuclear X-ray source in NGC~4395 was not detected
by the HRI.  We measure a 3~$\sigma$ upper limit of $1.0 \times 10^{-3}$
ct~s$^{-1}$ for the HRI count rate of this source.  Assuming the spectral
parameters obtained for the power-law model in the previous section, this
limit corresponds to a PSPC count rate of $2.9 \times 10^{-3}$ ct~s$^{-1}$,
which is even less than the source's ``low state'' during the PSPC observation.

We have examined the PSPC data to see if the count-rate variability of the
nuclear X-ray source in NGC~4395 was accompanied by spectral variations.  As
a crude means of investigating this issue, we have measured the ``hardness
ratio'' of the source in both segments of the PSPC observation.  The hardness
ratio is defined in the following manner: $HR = (H - S) / (H + S)$, where
$H$ and $S$ represent the number of source counts detected in ``hard''
(0.94--2.4 keV) and ``soft'' (0.1--0.94 keV) bands of the spectrum.  The
demarcation energy of 0.94 keV divides the {\it total\/} PSPC counts detected
from the nucleus equally into the two bands (\ie $HR = 0.0$ for the full
17~ks observation).  We measure $HR = -0.13 \pm 0.21$ for the first segment
of the observation, and $HR = +0.04 \pm 0.12$ for the second.  Although the
uncertainties are substantial, these results are consistent with there being
no spectral variability in conjunction with the intensity variations of the
nuclear X-ray source.

The issue of hardness ratio variability provides further insight into the
nature of the soft X-ray spectrum of NGC~4395.  There are two commonly
observed types of AGN spectra that might account for the S-shaped pattern
in the residuals seen in Figure~2.  The first of these consists of two
emission components: a thermal component that dominates at the softest
X-ray energies, and a moderately absorbed nonthermal component that
contributes mainly at higher energies.  Spectra of this form have been
successfully used to model the broadband X-ray emission of other low-luminosity
AGNs (Serlemitsos, Ptak, \& Yaqoob 1996).  The second type of model
consists of a single power-law component that is absorbed by both neutral
and ionized material.  The signatures of ionized (``warm'') absorption
are often observed in the spectra of luminous type~1 Seyferts (\eg Reynolds
1997).  In the two-component model, the thermal emission is presumed to
be spatially extended, so it should not vary on short time scales.  Thus,
in this picture, we might expect strong hardness ratio variability to
accompany the intensity variations observed in NGC~4395.  The absence of
significant changes in the hardness of the spectrum suggests instead that
only one emission component is present, which favors the warm absorber
scenario.

\subsec {2.2.} {Ultraviolet and Optical Spectroscopy}

Our investigation of the spectral energy distribution of NGC~4395 includes
published ultraviolet and optical spectra from Filippenko \et (1993) and
Kraemer \et (1999), which provide a full description of the acquisition,
reduction, and analyses of the data.  The salient details of these spectra
are summarized as follows.  The UV spectrum was acquired with the Faint
Object Spectrograph on board the {\sl Hubble Space Telescope}
(pre-refurbishment) through an effective aperture of 1\farcs4 $\times$
4\farcs3; it covers the 1150--3300~\AA\ range at a resolution of 1.2--3.3~\AA.
The ground-based optical spectrum, which covers the 3200--10000~\AA\ range
at a resolution of 5--8~\AA, was acquired under photometric conditions with
$\sim$~$1''$ seeing and a $2'' \times 4''$ aperture.

Extinction by dust in NGC~4395, if present, could affect the shape of the
galaxy's optical/UV spectrum and its apparent luminosity.  Because NGC~4395
is a face-on, bulgeless, low-metallicity spiral, extinction in the nucleus is
expected to be very low.  This is, in fact, what both optical emission-line
flux ratio measurements (Ho, Filippenko, \& Sargent 1997) and our power-law
fit to the {\sl ROSAT\/} PSPC spectrum (\S~2.1.1) suggest.  But this evidence
should be interpreted cautiously: the emission-line flux ratios reflect the
amount of extinction between us and the galaxy's narrow-line region, not its
continuum source, and as we have noted, a power law may not describe the
spectrum of the nuclear X-ray emission accurately.  As discussed in \S~3,
there is indirect evidence that the far-UV continuum of NGC~4395 is absorbed.
In addition, our UV spectrum exhibits an abrupt inflection near 2200~\AA\
(Filippenko \et 1993); this is approximately coincident with the strong
2175~\AA\ bump in the Galactic reddening law, which could indicate the
presence of dust.  In an attempt to constrain the extinction in NGC~4395,
we have applied the Galactic reddening law of Cardelli, Clayton, \& Mathis
(1989) using a range of values for the equivalent $V$-band absorption $A_V$.
For $A_V$ in excess of 0.2 mag, we find that the 2175~\AA\ feature is
over-corrected.  Although this procedure for constraining $A_V$ can be very
uncertain (see Fitzpatrick 1999), it nonetheless provides a practical upper
limit to the amount of extinction in the nucleus of NGC~4395.  But since
there is no compelling evidence for extinction intrinsic to NGC~4395,
we correct only for Galactic reddening when constructing the SED.

It has been recently reported that the shape of the optical continuum of
NGC~4395 undergoes dramatic changes on time scales as short as 6 months
(Lira \et 1999).  As we have been monitoring NGC~4395 for many years, we
are able to investigate this claim independently.  Figure~3 shows five
of our broadband optical spectra, which were acquired and reduced in the
same manner as the data described above.  Accompanying each spectrum is a
set of dotted lines that approximate the three ``states'' observed by Lira
\et (1999) over the same wavelength range.  As the dates listed in Figure~3
indicate, our spectra cover a variety of time baselines ranging from 2 months
to almost 4 years.  Qualitatively, the spectra all have the same general
appearance: none can be described as a power law, and all are concave
downward.  Although there appear to be some differences between the spectra,
we do not see evidence for the extreme spectral variability reported by Lira
\et (1999).  Moreover, the differences we observe may be caused, in part,
by minor inconsistencies in our flux calibration and extinction correction
procedures.  Unfiltered images of NGC~4395 obtained at five epochs spanning
40 days with the Katzman Automatic Imaging Telescope (KAIT; Treffers \et
1997) do indicate that the nucleus is variable by at least 0.25~mag.  Thus,
we have initiated a {\sl BVRI\/} monitoring campaign in order to
search for color variability.  We note, however, that the ``low state''
observed by Lira \et (1999), if real, cannot be very common in NGC~4395;
otherwise, the UV/optical spectrum would be characterized by low-ionization
emission lines, such as those in LINERs (Ho, Filippenko, \& Sargent 1993),
rather than the high-ionization lines that are consistently observed in the
spectrum.  Perhaps the ``low state'' spectrum was adversely affected by
non-grey particles in the Saharan dust mentioned by Lira et al.

\subsec {2.3.} {Infrared Observations}

Infrared images of the NGC~4395 nucleus were obtained using the Near-Infrared
Camera (Matthews \& Soifer 1994) on the Keck-I 10~m telescope on 1995 January
20 UT.  Five exposures with integration times of 45~s, 25~s, and 25~s were
acquired in each of the $J$, $H$, and $K$ bandpasses, respectively.  The
seeing during these observations ranged from approximately 0\farcs5 (FWHM) in
$K$ to 0\farcs7 in $J$. Fluxes were measured within a 1\farcs5 radius aperture,
and the standard star G163$-$50 (Casali \& Hawarden 1992) was used for
absolute flux calibration.  The following magnitudes and 1~$\sigma$ errors
were derived: $J = 15.58 \pm 0.06$~mag, $H = 14.98 \pm 0.05$~mag, and
$K = 14.35 \pm 0.06$~mag.  While these magnitudes are dominated by the
nuclear point source, some very low-level contamination from faint
circumnuclear starlight may be present.

We have measured the strength of the continuum of NGC~4395 at 3.94~$\mu$m,
based on an 18.6~ks observation with the Short Wavelength Spectrometer (SWS)
on board the {\sl Infrared Space Observatory}$^9$.  Full details of the
observation and data reduction are provided in Kraemer \et (1999).  The
3.9~$\mu$m flux density of NGC~4395 is $6.7 \pm 0.7$~mJy, after removal of
the estimated background of 0.7~mJy due to zodiacal emission.  Based on
analysis of our $K$-band image, we estimate further that $\sim$~30\% of
this flux arises from extranuclear sources in the $14'' \times 20''$ SWS
aperture; applying this correction, we derive a 3.9~$\mu$m flux density
of $4.7 \pm 0.7$~mJy for the nucleus.
\vfootnote{$^9$}{The {\sl Infrared Space Observatory\/} is an ESA project
with instruments funded by ESA Member States with the participation of
ISAS and NASA.\vskip -10pt}
\vfootnote{$^{10}$} {The National Radio Astronomy Observatory is a facility
of the National Science Foundation operated under cooperative agreement by
Associated Universities, Inc.}

NGC~4395 has also been observed in the $N$ band by Maiolino \et (1995),
who report a 10~$\mu$m flux density of $12 \pm 3.9$~mJy.  A 5\farcs3
diameter aperture was employed in this observation, so a small contribution
from the galaxy may be present in this measurement as well.

\subsec {2.4.} {Radio Observations}

Radio continuum observations of NGC~4395 were made at a wavelength of 20~cm
on 1990 March~3 with the NRAO$^{10}$ Very Large Array (VLA), which was operated
in the high-resolution A~configuration.  The on-source integration was about
eight hours.  Phase calibration was performed using the nearby unresolved
source 1219+285, and 3C~286 was used to establish the amplitude scale.  The
rms noise level in the 20~cm image, which was presented previously by Sramek
(1992), is 0.036 mJy.  We also analyzed an archived 6~cm VLA A-array
observation of NGC~4395 from 1982 February~8.  The rms noise level in the
6~cm map is 0.12 mJy.  Both radio images were convolved to a beam size of
2\farcs7; at this resolution, the nuclear radio source in NGC~4395 is
spatially unresolved.  After applying corrections for primary beam attenuation
(0.87 at 6~cm) and attenuation due to bandwidth effects (0.84 at 20~cm and
0.90 at 6~cm), we obtain the following source flux densities: $S_{20} =
1.24 \pm 0.07$~mJy and $S_6 = 0.56 \pm 0.12$~mJy.  The uncertainties include
a 5\% calibration error contribution.  For a power-law spectrum of the form
$S_{\nu} \propto \nu^{-\alpha}$, these flux densities imply a radio spectral
index$^{11}$ of $\alpha = 0.66 \pm 0.22$.
\vfootnote{$^{11}$} {Our spectral index measurement assumes that the nuclear
radio source in NGC~4395 did not vary between the 6~cm and 20~cm observations.
NGC~4395 was detected in the VLA FIRST survey (White \et 1997) with a 20~cm
flux density of 1.2~mJy, equal to the value we obtained in 1990, suggesting
that this assumption is valid.\vskip -10pt}

\sec {3.} {The Nuclear SED of NGC 4395}

\fpsec
We have combined the data described in the previous section to construct the
observed spectral energy distribution of NGC~4395.  The result, plotted in
$\nu L_{\nu}$ units and corrected for Galactic absorption, is displayed in
Figure~4$a$.  The X-ray region of the SED (0.2--2~keV) is depicted as the
power law discussed in \S~2.1.1, normalized to be consistent with the average
PSPC count rate.  Again, this power law provides a reasonable approximation
of the observed soft X-ray spectrum, but it is unlikely to be accurate in
detail.  The dotted line between $3.3 \times 10^{15}$ Hz and $4.8 \times
10^{16}$ Hz (i.e., 13.6~eV and 200~eV) was not measured; we use it below to
estimate the ionizing photon luminosity of the nucleus.

We are confident that Figure~4$a$ represents the {\it nuclear\/} SED of
NGC~4395, since it consists mainly of data obtained with resolutions and/or
apertures a few arcseconds or less in extent.  (For a distance of 2.6~Mpc,
$1''$ corresponds to 13~pc.)  The two exceptions---the {\sl ISO\/} and
{\sl ROSAT\/} measurements---are also very likely to reflect nuclear fluxes.
As discussed in \S~2.3, we have corrected the {\sl ISO\/} 3.9~$\mu$m flux
density for extranuclear contamination.  Furthermore, based on our comparison
of the {\sl ROSAT\/} PSPC, {\sl ROSAT\/} HRI, and optical images of NGC~4395,
we have shown that the source of soft X-ray emission identified as source~A
in Figure~1 is very likely to be associated with the galaxy's nucleus.
The amplitude and time scale of the 0.2--2~keV variability implies that the
majority of the nuclear soft X-ray emission arises from a very compact
region ($r < 0.01$ pc), leaving little room for a significant extranuclear
component.

As Figure 4$a$ indicates, the nuclear SED of NGC~4395 peaks at least twice,
once somewhere in the near-IR band, and again in the optical band at
$\sim$~$5 \times 10^{14}$ Hz ($\sim$~6000~\AA ).  The SED is also rising
in the UV above $\sim$~$2 \times 10^{15}$ Hz (which is the case even if an
extinction correction of $A_V = 0.2$ mag is applied).  Integration of the
SED yields a bolometric luminosity of $L_{\rm bol} \approx$ \EE {1.9} {40}
for the nucleus of NGC~4395 (\EE {3.2} {40} for $A_V = 0.2$ mag), close to
the value of \EE {1.5} {40} predicted by Filippenko \& Sargent (1989) on
the basis of the observed broad \Hb\ line luminosity.
\vfootnote{$^{12}$}{We have assumed that the continuum can be described as
simple power laws over the ranges 5~GHz to 10~$\mu$m, 10~$\mu$m to 3.9~$\mu$m,
3.9~$\mu$m to 2.2~$\mu$m, and 13.6~eV to 200~eV.}

We have used the artificial power law connecting the observed ultraviolet
spectrum (extrapolated to 13.6 eV) and the time-averaged soft X-ray spectrum
to obtain a conservative estimate of the nuclear ionizing photon luminosity
$Q$.  Integration of the SED between 13.6~eV and 2~keV yields $Q_{\rm obs} =
3.1 \times 10^{49}$ photons~s$^{-1}$ (= $5.3 \times 10^{49}$ photons~s$^{-1}$
if $A_V = 0.2$ mag).  For Case~B recombination (Osterbrock 1989), the
ionizing photon luminosity required to account for the total observed
\Hb\ luminosity $L$(\Hb) $= 4.4 \times 10^{37}$ ergs s$^{-1}$ (Filippenko
\& Sargent 1989) is $Q_{\rm req} = 9.2 \times 10^{49}\> f^{-1}$ photons
s$^{-1}$, where $f$ is the combined covering factor of the broad and narrow
emission-line gas.  Although $f$ is unmeasured, it must have a value less
than unity; thus, $Q_{\rm obs} < Q_{\rm req}$, which implies a deficit of
ionizing radiation for this form of the SED.  Since the nuclear X-ray source
in NGC~4395 is variable, the galaxy's average ionizing luminosity may be
higher than we have estimated it to be.  On the other hand, it is unlikely
that the fiducial far-UV spectrum we have adopted describes the correct
shape of the ionizing continuum.  As noted above, the UV portion of the
observed SED is rising (in $\nu L_{\nu}$) at frequencies greater than
$\sim$~$2 \times 10^{15}$~Hz.  Therefore, the actual far-UV spectrum of
NGC~4395 probably contains flux in excess of the power law we have assumed.
Kraemer \et (1999) have noted that this power law, which has an energy
index of $\alpha \approx 3$ for $F_{\nu} \propto \nu^{-\alpha}$, is far
too steep to account for the observed \HeII /\Hb\ emission-line flux ratio.
Interestingly, the slope $\alpha = 1.7$ needed to explain the \HeII /\Hb\
ratio would also be sufficient to eliminate the apparent deficit of $Q$ for
$A_V = 0.2$ mag.  In any case, the intrinsic 13.6--200~eV spectrum (or a
part of it) is almost certainly flatter than indicated in Figure~4$a$,
which may signify that the absorption column density in the nucleus of
NGC~4395 is larger than suggested by our power-law fit to the PSPC
spectrum.

A comparison between the nuclear SED of NGC~4395 and the median SEDs of
radio-quiet quasars (compiled by Elvis \et 1994) and typical low-luminosity
AGNs (LLAGNs, compiled by Ho 1999a) is provided in Figures 4$b$ and 4$c$,
respectively.  In each plot, the reference SED is shown as a dotted line
arbitrarily normalized to the SED of NGC~4395 at $1.36 \times 10^{14}$ Hz
(2.2~$\mu$m).  For clarity,
we have replaced the optical/UV spectrum of NGC~4395 with a continuum fit.
The SED of NGC~4395, particularly from the near-IR to the UV, differs
dramatically from the SEDs of both quasars and LLAGNs (which, as noted by
Ho 1999b, are very different from each other).  In NGC~4395, the SED rises
in the near-IR and declines in the UV, whereas the quasar SED has a local
minimum in the near-IR and rises in the UV.  None of this structure is
present in the median LLAGN SED, which declines steadily from the near-IR
to the far-UV.  Relative to the flux density at 2.2~$\mu$m, NGC~4395 is a
stronger radio source than the typical radio-quiet quasar, but it is not
as radio-loud as other LLAGNs.  In the far-UV and soft X-ray region, the
SED of NGC~4395 also appears to deviate from the quasar and LLAGN SEDs;
however, further characterization of the X-ray spectrum and variability of
NGC~4395 is required before this conclusion can be drawn.

At this time we are unable to provide a full explanation for the differences
between the SEDs of quasars, typical LLAGNs, and NGC~4395, but we offer the
following qualitative argument as a starting point for an investigation of
the issue. As Blandford \& Rees (1992) have discussed, the radiation energy
density of an optically thick region a fixed distance (in gravitational radii)
from a massive, collapsed object depends only on $M_{\rm BH}$, the mass of the
object (assumed here to be a black hole), and $L_{\rm bol} / L_{\rm edd}$,
the ratio of the bolometric to Eddington luminosity of the region.  It would
therefore be of interest to compare $M_{\rm BH}$ and $L_{\rm bol} /
L_{\rm edd}$ for quasars, LLAGNs, and NGC~4395.  For typical high-luminosity
quasars, it is presumed that $M_{\rm BH} > 10^9$~\Msun\ and $L_{\rm bol} /
L_{\rm edd} \approx$ 10$^{-1}$--10$^{-2}$ (\eg Padovani \& Rafanelli 1988;
McLure \et 1999).  The majority of the LLAGNs from Ho (1999a) have $M_{\rm BH}
\approx$ $10^6$--$10^9$~\Msun\ and $L_{\rm bol} / L_{\rm edd} \approx$
10$^{-4}$--10$^{-6}$.  In the case of NGC~4395, the stellar kinematics near
the nucleus place an upper limit of $8 \times 10^4$~\Msun\ for the mass of
the central black hole (Filippenko \& Ho 1999), which implies $L_{\rm edd}
<$ \EE {1} {43} and $L_{\rm bol} / L_{\rm edd} > 2 \times 10^{-3}$.  The
latter limit will be even higher if there is significant extinction in the
nucleus of NGC~4395.  Clearly, quasars, LLAGNs, and NGC~4395 occupy very
different regions in the $M_{\rm BH}$ -- $L_{\rm bol} / L_{\rm edd}$ plane,
which may be partly responsible for the dissimilarity of their SEDs.  Of
course, there are undoubtedly other important factors to consider.  For
example, most of the LLAGNs studied by Ho (1999a) are LINERs, which, unlike
quasars and luminous Seyfert galaxies, may be powered by advection-dominated
accretion flows rather than geometrically thin accretion disks (\eg Lasota
\et 1996).  Further investigation of the broadband characteristics of active
galaxies, particularly those with measured black hole masses, will clarify
the relationship between their physical properties and their SEDs.

\sec {4.} {Nature of the Nuclear X-ray Source}

\fpsec
The small physical size of the nuclear X-ray source in NGC~4395, deduced
from its large-amplitude (300\%), short time scale ($\sim$~10~d) variability,
suggests that this emission results from the accretion of matter onto a
massive, compact object.  But with a mean observed 0.2--2 keV luminosity
of just \EE {1.3} {38} (calculated for the $\alpha = -0.1$ power-law spectrum
described in \S~2.1.1 and a distance of 2.6~Mpc), the physical nature of the
accreting object is somewhat unclear.  While an AGN could easily account for
this modest luminosity, it could also be produced by a single bright binary
star system emitting near the Eddington limit for a stellar-mass neutron
star or black hole.  Some of the luminous X-ray emitters identified recently
in the off-nuclear regions of nearby galaxies might be examples of the
latter (Colbert \& Mushotzky 1999).  Both AGNs and X-ray binaries are known
to flicker, so the variability of the source does not help to distinguish
between the two possibilities.  The nuclear X-ray spectrum of NGC~4395,
because of its apparent flatness, bears a closer resemblance to the spectra
of high-mass X-ray binaries ($\bar{\alpha} \approx 0.2$; White, Swank, \&
Holt 1983) than to those of luminous broad-line AGNs ($\bar{\alpha} \approx
1.5$ in the {\sl ROSAT\/} band; Walter \& Fink 1993), which might be cited
as support for the binary hypothesis.  However, as we have discussed above,
this spectrum probably possesses complex features, so conclusions regarding
its true shape must be deferred until high-resolution broadband observations
from {\sl Chandra\/} become available.  High-mass X-ray binaries emit most
of their luminosity in the X-ray band, so if the nuclear X-ray source were
a binary star system, the slope of the galaxy's far-UV spectrum would have
to be even {\it steeper\/} than it is shown to be in Figure~4$a$. This
circumstance is firmly ruled out by the observed strength of the \HeII /\Hb\
optical emission-line flux ratio (Kraemer \et 1999).  Thus, despite the
scarcity of direct evidence linking the nuclear X-ray emission in NGC~4395
to an AGN, we feel that it is very likely to be associated with the object
responsible for the galaxy's optical and UV emission-line properties.  Some
of the off-nuclear X-ray sources shown in Figure~1, however, could be luminous
X-ray binaries or supernova remnants, as discussed by Colbert \& Mushotzky
(1999).

\sec {5.} {Summary}

\fpsec
We have presented the nuclear radio--to--X-ray spectral energy distribution
of NGC 4395, the least luminous broad-line Seyfert galaxy currently known.
While this object exhibits many of the characteristics of its more luminous
counterparts, its SED differs markedly.  Interestingly, the SED of NGC~4395
does not resemble the SEDs of other (predominantly LINER) low-luminosity
AGNs either, which themselves differ from the typical quasar SED.  The
full reason for the dissimilarity between these SEDs is uncertain.  We have
noted, however, that NGC~4395, quasars, and LLAGNs have different combinations
of black hole mass and Eddington-normalized bolometric luminosity, which
may be a significant factor.  Continued broadband study of AGNs of all
spectroscopic classes and luminosities is needed to elucidate the connection
between the shape of their SEDs and their physical attributes.

The {\sl ROSAT\/} data discussed here represent the first detection of X-ray
emission from the nucleus of NGC~4395.  Although weak, the nuclear X-ray
source exhibits large-amplitude variability on short time scales and has an
apparently flat soft X-ray spectrum.  A simple absorbed power-law model,
however, does not provide an acceptable fit to this spectrum; we speculate
that the spectrum may contain additional emission or absorption components. 
There may be some ambiguity about the origin of the nuclear X-ray emission
because of its faintness, but we have argued that if it were associated with
anything other than an AGN, it would be very difficult (or impossible) for
NGC~4395 to power the Seyfert-like emission lines observed in the optical
and UV portions of its spectrum.  {\sl Chandra\/} observations should clarify
the nature of this source.

\medskip
We would like to express our thanks to Kirk Korista, the referee, for his
thoughtful comments on the manuscript.  We are also grateful to M.~Eracleous
for helpful discussions, to D.~DePoy, J.~Graham, W.~Harrison, and M.~Liu for
assistance with the Keck observations and data analysis, and to W.~Li and
M.~Papenkova for providing the KAIT magnitude measurements.  Support for this
work was provided by NASA through grants NAG5-3556, NAG5-3563, NAG5-3690,
and AR-07527 (through STScI, which is operated by AURA, Inc., under NASA
contract NAS5-26555).  ECM acknowledges partial support by NASA through 
Chandra Fellowship grant PF8-10004 awarded by the Chandra X-ray Center,
which is operated by the Smithsonian Astrophysical Observatory for NASA
under contract NAS8-39073. The W.~M.~Keck Observatory is operated as a
scientific partnership between the California Institute of Technology, the
University of California, and NASA, and was made possible by the generous
financial support of the W.~M.~Keck Foundation.

\break
\centerline{\apj References}

\bigskip

{\baselineskip 12pt\smrm
\hi Blandford, R.\ D., \& Rees, M.\ J.\ 1992, in Testing the AGN Paradigm,
eds.\ S.\ S. Holt, S.\ G.\ Neff, \& C.\ M.\ Urry (New York:~AIP), 3

\hi Cardelli, J.\ A., Clayton, G.\ C., \& Mathis, J.\ S.\ 1989, ApJ, 345, 245

\hi Casali, M.\ M., \& Hawarden, T.\ G.\ 1992, JCMT-UKIRT Newsletter, No.~3, 33

\hi Colbert, E.\ J.\ M., \& Mushotzky, R.\ F.\ 1999, ApJ, in press

\hi Elvis, M., et al.\ 1994, ApJS, 95, 1

\hi Filippenko, A.\ V., \& Ho, L.\ C.\ 1999, ApJ, submitted

\hi Filippenko, A.\ V., Ho, L.\ C., \& Sargent, W.\ L.\ W.\ 1993, ApJ, 410, L75

\hi Filippenko, A.\ V., \& Sargent, W.\ L.\ W.\ 1989, ApJ, 342, L11

\hi Fitzpatrick, E.\ L.\ 1999, PASP, 111, 63

\hi Hasinger, G., Turner, T.\ J., George, I.\ M., \& Boese, G.\ 1992, Legacy,
2, 77

\hi Ho, L.\ C.\ 1999a, ApJ, in press

\hi Ho, L.~C.\ 1999b, in  The AGN-Galaxy Connection, eds.\ H.~R.\ Schmitt,
L.~C.\ Ho, \& A.~L.\ Kinney (Adv.\ Space Res.), in press (astro-ph/9807273)

\hi Ho, L.\ C., Filippenko, A.\ V., \& Sargent, W.\ L.\ W.\ 1993, ApJ, 417, 63

\hi Ho, L.\ C., Filippenko, A.\ V., \& Sargent, W.\ L.\ W.\ 1997, ApJS, 112, 315

\hi Kraemer, S.\ B., Ho, L.\ C., Crenshaw, D.\ M., Shields, J.\ C., \&
Filippenko, A.\ V.\ 1999, ApJ, in press

\hi Lasota, J.-P., Abramowicz, M.~A., Chen, X., Krolik, J., Narayan, R., \&
Yi, I.\ 1996, ApJ, 462, 142

\hi Lira, P., Lawrence, A., O'Brien, P., Johnson, R.\ A., Terlevich, R., \&
Bannister, N.\ 1999, MNRAS, in press

\hi Maiolino, R., Ruiz, M., Rieke, G.\ H., \& Keller, L.\ D.\ 1995, ApJ, 446,
561

\hi Matthews, K., \& Soifer, B.\ T.\ 1994, in Infrared Astronomy with Arrays,
The Next Generation, ed.\ I.\ McLean (Dordrecht:~Kluwer), 239

\hi McCall, M.\ L., Rybski, P.\ M., \& Shields, G.\ A.\ 1985, ApJS, 57, 1

\hi McLure, R.\ J., Dunlop, J.\ S., Kukula, M.\ J., Baum, S.\ A., O'Dea,
C.\ P., \& Hughes, D.\ J.\ 1999, ApJ, in press

\hi Murphy, E.\ M., Lockman, F.\ J., Laor, A., \& Elvis, M.\ 1996, ApJS, 105,
365

\hi Osterbrock, D.\ E.\ 1989, Astrophysics of Gaseous Nebulae and Active
Galactic Nuclei, (University Science Books: Mill Valley)

\hi Padovani, P., \& Rafanelli, P.\ 1988, A\&A, 205, 53

\hi Pfeffermann, E., et al.\ 1986, SPIE, 733, 519

\hi Reynolds, C.\ S.\ 1997, MNRAS, 286, 513

\hi Serlemitsos, P., Ptak, A., \& Yaqoob, T.\ 1996, in The Physics of LINERs in
View of Recent Observations, eds.\ M.\ Eracleous, A.\ Koratkar, C.\ Leitherer,
\& L.\ Ho, (San Francisco: ASP Conf.\ Ser.\ Vol.\ 103), p.~70

\hi Sramek, R.\ 1992, in Relationships Between Active Galactic Nuclei and
Starburst Galaxies, ed.\ A.\ V.\ Filippenko (San Francisco:~ASP Conf.\ Ser.\
Vol.\ 31), 273

\hi Treffers, R.~R., Peng, C.~Y., Filippenko, A.~V., \& Richmond, M.~W.\ 1997, IAU Circ.\ No.\ 6627.

\hi Tr\"umper, J. 1983, Adv.\ Space Res., 2(4), 241

\hi Walter, R., \& Fink, H.\ H.\ 1993, A\&A, 274, 105

\hi White, N., Swank, J., \& Holt, S.\ 1983, ApJ, 270, 711

\hi White, R.\ L., Becker, R.\ H., Helfand, D.\ J., \& Gregg, M.\ D.\ 1997,
ApJ, 475, 479
\par}

\break
\centerline{\apj Figure Captions}

\fpsec\noindent
{\apj Fig.~1.}---Contours from the {\sl ROSAT\/} PSPC image of NGC~4395
overlayed on an $8' \times 8'$ optical image from the Digitized Sky Survey.
The PSPC image was smoothed with a $\sigma_{\rm G} = 7''$ Gaussian.  The
contours represent intensity levels of 2, 3, 4, 5, 9, 16, and 31 (smoothed)
counts per pixel.  Source~A is coincident with the optical nucleus of
NGC~4395.  None of the other sources detected appears to correspond to an
optical feature.  As a result of the smoothing, source~B appears to be
partially merged with the nuclear X-ray source; the two sources are well
resolved in the raw X-ray image.  The ordinate (declination) and abscissa
(right ascension) are labeled with J2000 coordinates.

\bigskip\noindent
{\apj Fig.~2.}---The {\sl ROSAT\/} PSPC counts spectrum of NGC~4395, fitted
with the absorbed power-law model described in the text.  The systematic,
rather than random, displacement of the fit residuals suggests that the
spectrum possesses complex features.

\bigskip\noindent
{\apj Fig.~3.}---Five optical spectra of NGC~4395, obtained over a $\sim$~4
year period at Lick Observatory.  The strong emission lines in the spectra
have been truncated to emphasize the shape of the continuum.  The dotted
lines, normalized to intersect the spectrum of NGC~4395 near a wavelength
of 6300~\AA , approximate the various spectral states observed by Lira \et
(1999) over the same wavelength range.  While there are some differences
among our spectra, we do not see evidence for the extreme spectral
variability claimed by Lira \et (1999).

\bigskip\noindent
{\apj Fig.~4.}---($a$) The nuclear radio--to--X-ray spectral energy
distribution of NGC~4395, corrected for Galactic extinction ($A_V$ = 0.02
mag).  We have not attempted to correct for absorption that may be intrinsic
to NGC~4395.  As described in the text, we have artificially added the far-UV
power law ({\it dotted line}) between 13.6~eV and 200~eV.
($b$) Comparison of the SED of NGC~4395 to the median radio-quiet quasar (RQQ)
SED ({\it dotted line}) from Elvis \et (1994).  For clarity, we have replaced
the spectrum of NGC~4395 in the optical/UV region with a continuum fit.  The
quasar SED has been normalized to intersect the SED of NGC~4395 at 2.2~$\mu$m.
($c$) The SED of NGC~4395 compared to the median SED of the low-luminosity
AGNs ({\it dotted line}) studied by Ho (1999a).  In constructing the median
LLAGN SED, we have omitted the optical/UV spectra of two objects (NGC~4261
and NGC~4374) that have heavy amounts of internal reddening.  Again, the
LLAGN SED has been normalized to intersect the SED of NGC~4395 at 2.2~$\mu$m.

\break
\topglue 1.00truein
{\epsfxsize=7.5truein
\epsffile[126 180 468 477]{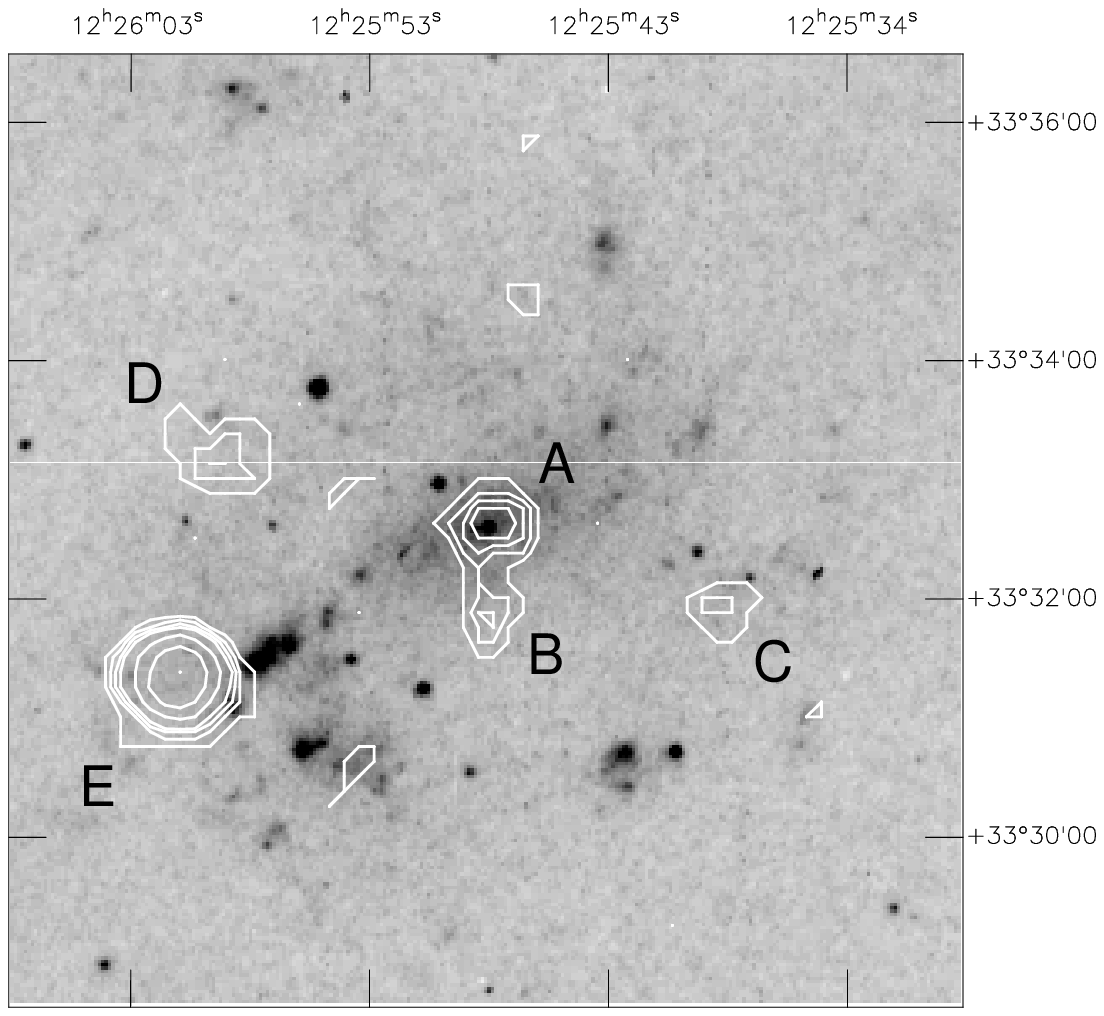}\par}

\vfill
\centerline{\apj Figure 1}

\break
\topglue 0.65truein
{\epsfxsize=6.0truein
\epsffile{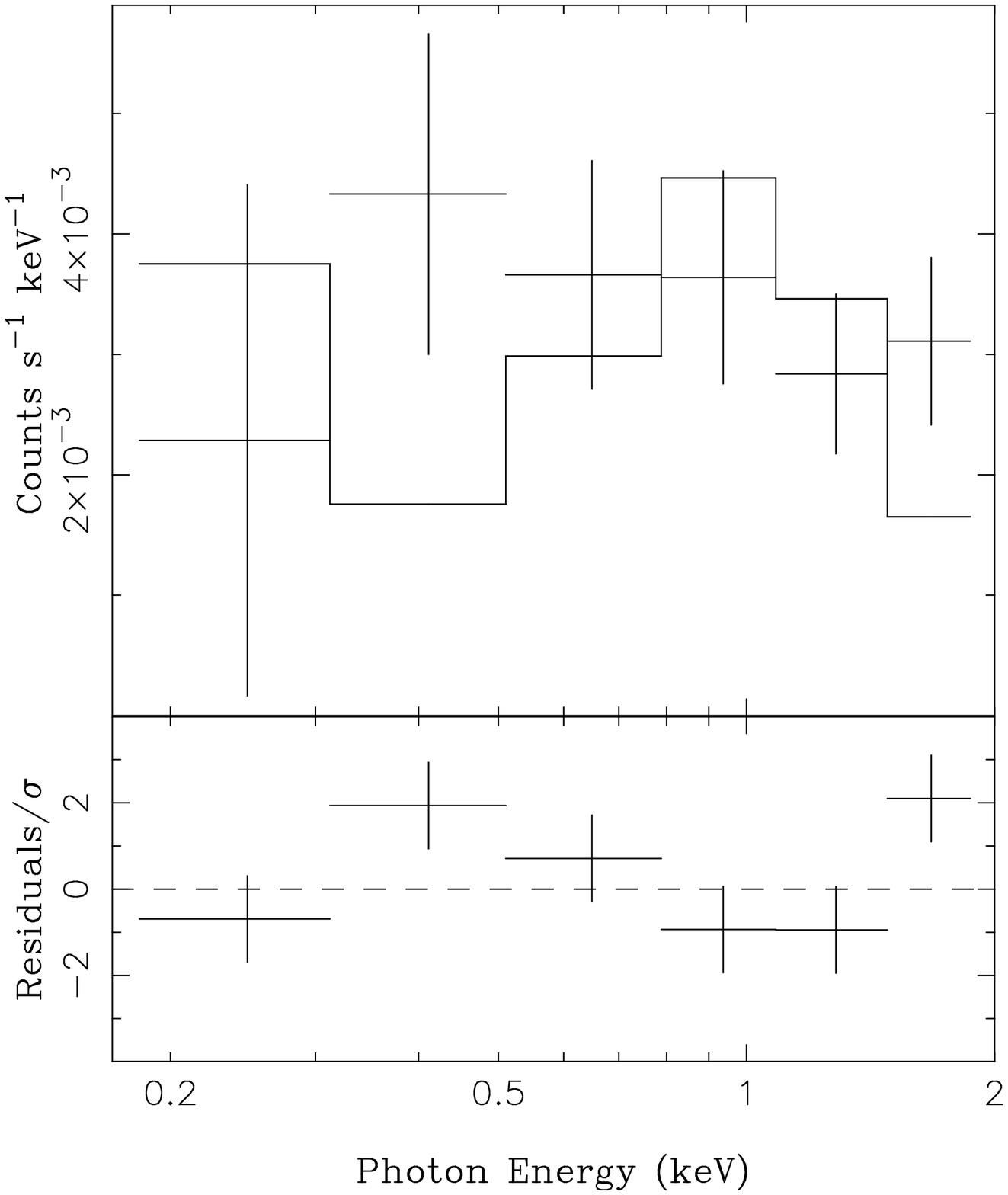}\par}

\vskip 0.9truein
\centerline{\apj Figure 2}

\break
\topglue 0.15truein
{\hskip -0.3truein
\epsfxsize=5.8truein
\epsffile{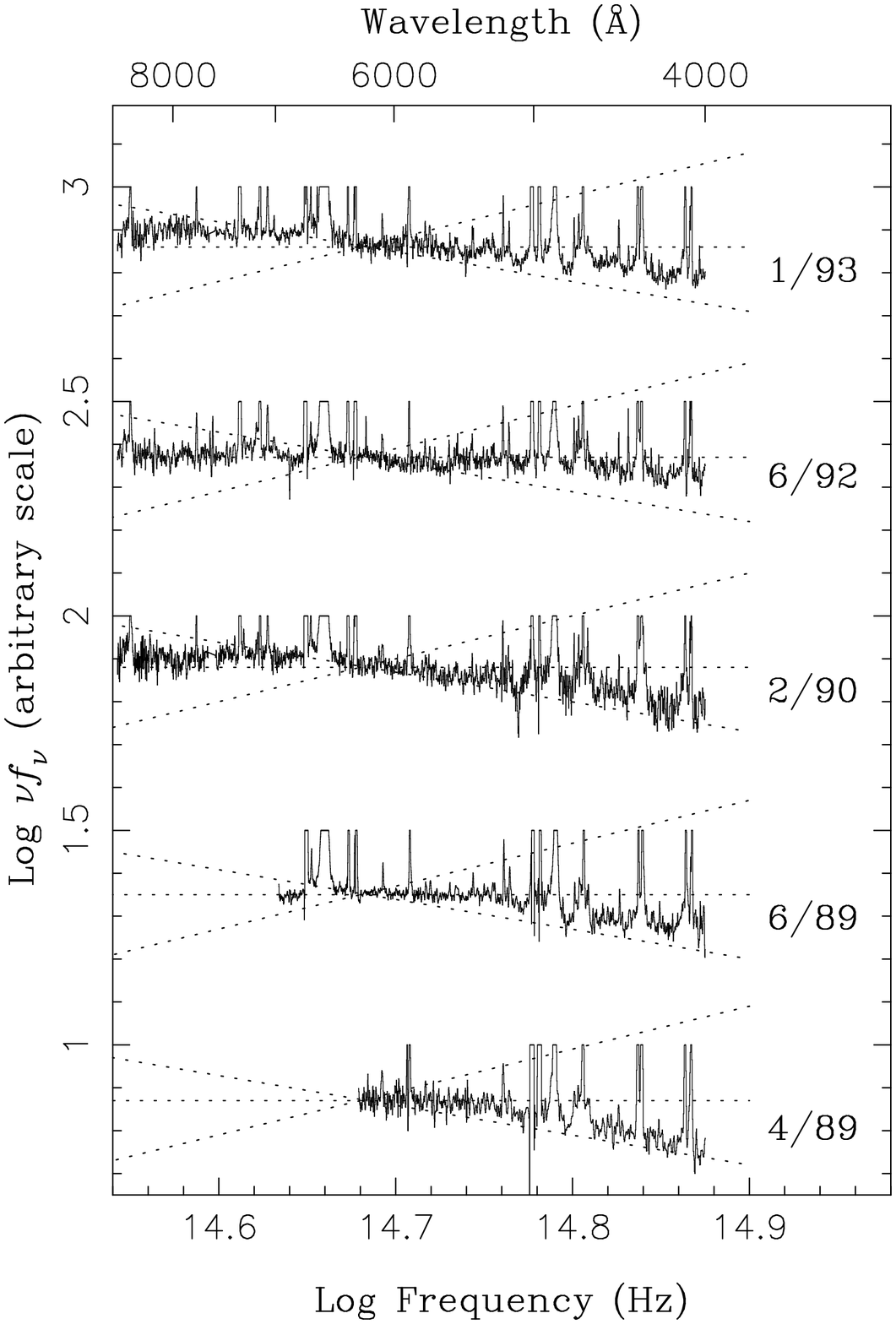}\par}

\vfill
\centerline{\apj Figure 3}

\break
{\hskip 0.00truein
\epsfxsize=5.1truein
\epsffile{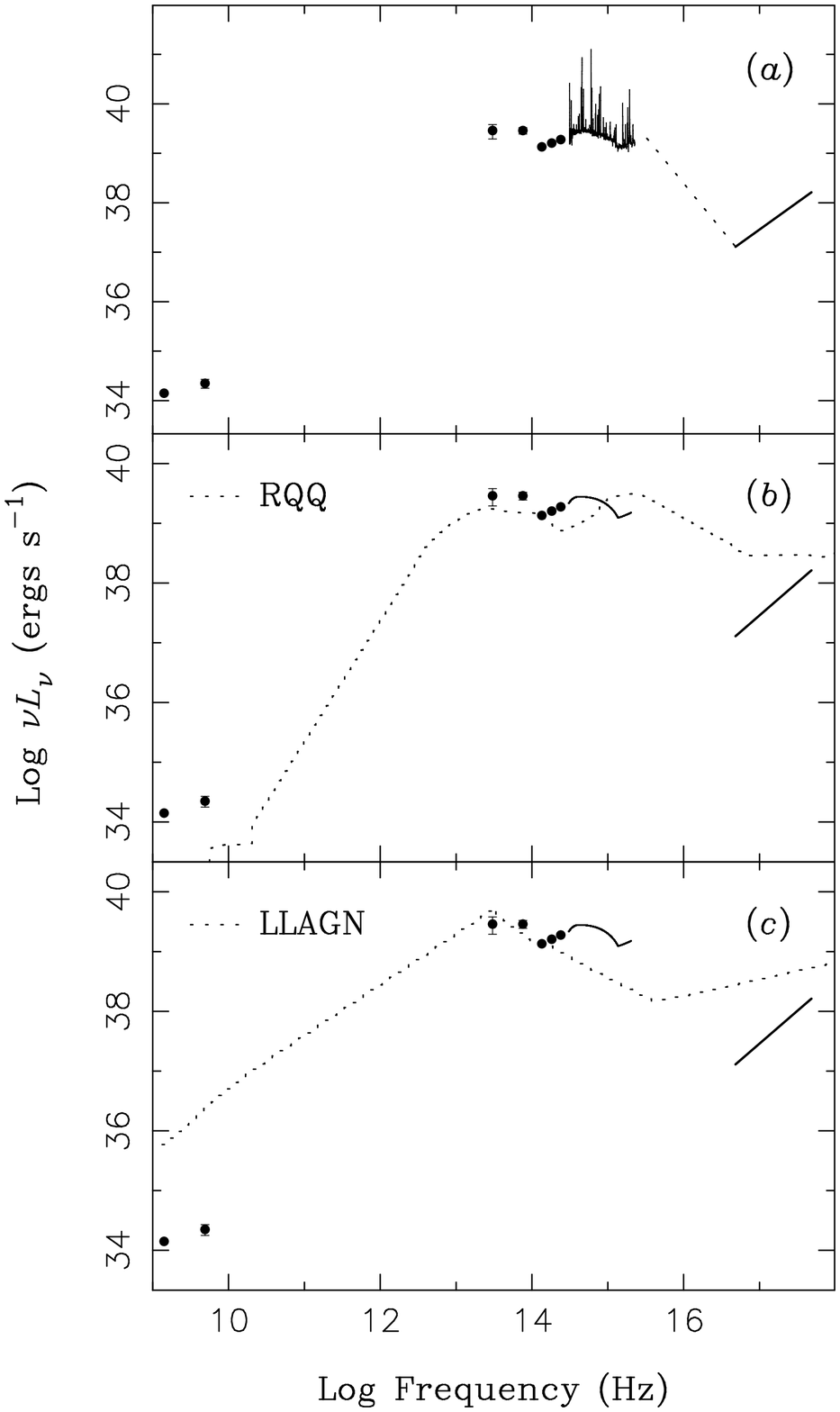}\par}

\vfill
\centerline{\apj Figure 4}

\bye